\RequirePackage{fix-cm}
\documentclass[smallcondensed]{svjour3}     
\smartqed  
\usepackage{graphicx}
\usepackage[utf8]{inputenc}
\usepackage{float}
\usepackage[english]{babel}
\usepackage{sidecap}
\usepackage{caption}
\usepackage{amsmath}
\usepackage{braket}
\usepackage{multirow}
\usepackage{physics}

\begin{document}

%
\title{Simulation studies of annihilation-photon's polarisation via Compton scattering with the J-PET tomograph}

\author{
N.~Krawczyk
\and
B.~C.~Hiesmayr
\and    
C.~Curceanu
\and
E.~Czerwi{\'n}ski
\and
K.~Dulski
\and
A.~Gajos
\and
M.~Gorgol
\and
N.~Gupta-Sharma
\and
B.~Jasi{\'n}ska
\and
D.~Kisielewska
\and
G.~Korcyl
\and
P.~Kowalski
\and   
W.~Krzemie{\'n}
\and
T.~Kozik
\and
E.~Kubicz
\and
M.~Mohammed
\and    
Sz.~Nied{\'z}wiecki
\and
M. Pa{\l}ka
\and
M.~Pawlik-Nied{\'z}wiecka
\and
L.~Raczy{\'n}ski
\and
J.~Raj
\and
K.~Rakoczy
\and
Z.~Rudy
\and
S.~Sharma
\and
Shivani
\and
R.Y.~Shopa
\and
M.~Silarski
\and
M.~Skurzok
\and
W.~Wi{\'s}licki
\and
B.~Zgardzi{\'n}ska
\and
M.~Zieli{\'n}ski
\and
P.~Moskal
}


\institute{N.~Krawczyk (corresponding author: nikodem.krawczyk@gmail.com) \and E.~Czerwi{\'n}ski \and K.~Dulski \and A.~Gajos \and N.~Gupta-Sharma \and K.~Kacprzak \and L.~Kaplon \and D.~Kisielewska \and G.~Korcyl \and T.~Kozik \and E.~Kubicz \and M.~Mohammed 
\and Sz.~Nied{\'z}wiecki \and 
M.~Pa{\l}ka \and M.~Pawlik-Nied{\'z}wiecka \and J. Raj \and K.~Rakoczy \and Z.~Rudy \and S.~Sharma \and Shivani \and M.~Silarski  \and M.~Skurzok \and M.~Zieli{\'n}ski \and P.~Moskal \at Faculty of Physics, Astronomy and Applied Computer Science, Jagiellonian University,  S.~Łojasiewicza 11, 30-348 Kraków, Poland\label{WFAIS}
    \and 
C.~Curceanu \at
    INFN, Laboratori Nazionali di Frascati CP 13,  Via E. Fermi 40, 00044, Frascati, Italy\label{LNF}
    \and
    B.~C.~Hiesmayr \at
    Faculty of Physics, University of Vienna  Boltzmanngasse 5, 1090 Vienna, Austria\label{Vienna}
    \and
    M.~Gorgol \and B.~Jasi{\'n}ska \and B.~Zgardzi{\'n}ska \at
    Department of Nuclear Methods, Institute of Physics, Maria Curie-Sklodowska University, Pl.~M.~Curie-Sklodowskiej~1, 20-031 Lublin, Poland\label{UMCS}
    \and
    P.~Kowalski \and L.~Raczy{\'n}ski \and R. Shopa \and W.~Wi{\'s}licki \at
    Department of Complex Systems, National Centre for Nuclear Research,  05-400 Otwock-Świerk, Poland\label{SWIERK}
    \and
    W.~Krzemie{\'n} \at
    High Energy Department, National Centre for Nuclear Research,  05-400 Otwock-Świerk, Poland\label{SWIERKHEP}
}
            
\maketitle

\date{Received: date / Accepted: date}


\begin{abstract}
J-PET is the first positron-emission tomograph (PET) constructed from plastic scintillators. It was optimized for the detection of photons from electron-positron annihilation. Such photons, having an energy of 511 keV, interact with electrons in plastic scintillators predominantly via the Compton effect. Compton scattering is at most probable at an angle orthogonal to the electric field vector of the interacting photon. Thus registration of multiple photon scatterings with J-PET enables to determine the polarization of the  annihilation photons. In this contribution we present estimates on the physical limitation in the accuracy of the polarization determination of $511$~keV photons with the J-PET detector.
\keywords{Compton Scattering; J-PET; Polarization}
\end{abstract}

\section{Introduction}
\begin{figure}[H]
\centering
 \includegraphics[width=0.49\textwidth]{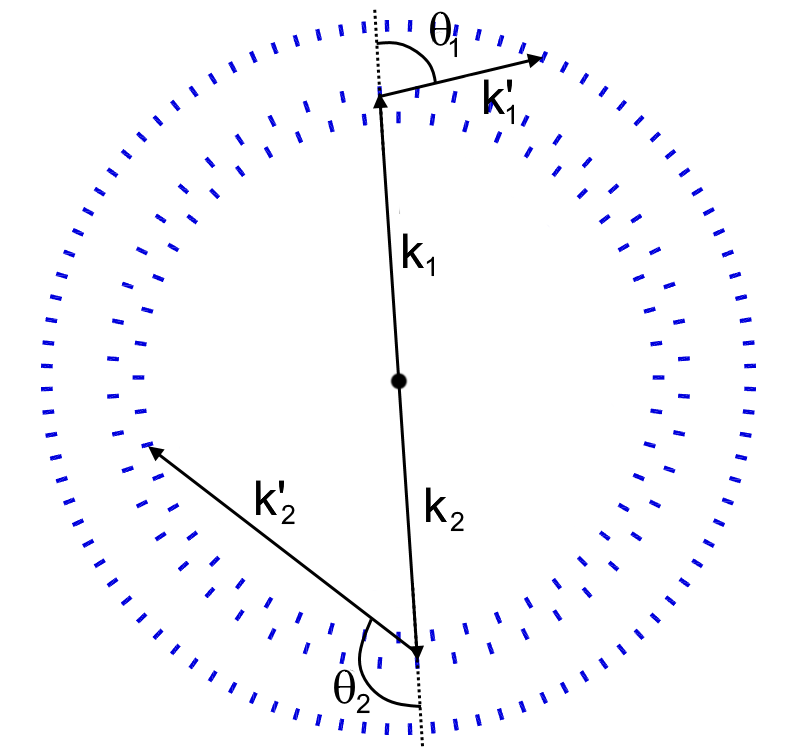}
  \includegraphics[width=0.49\textwidth]{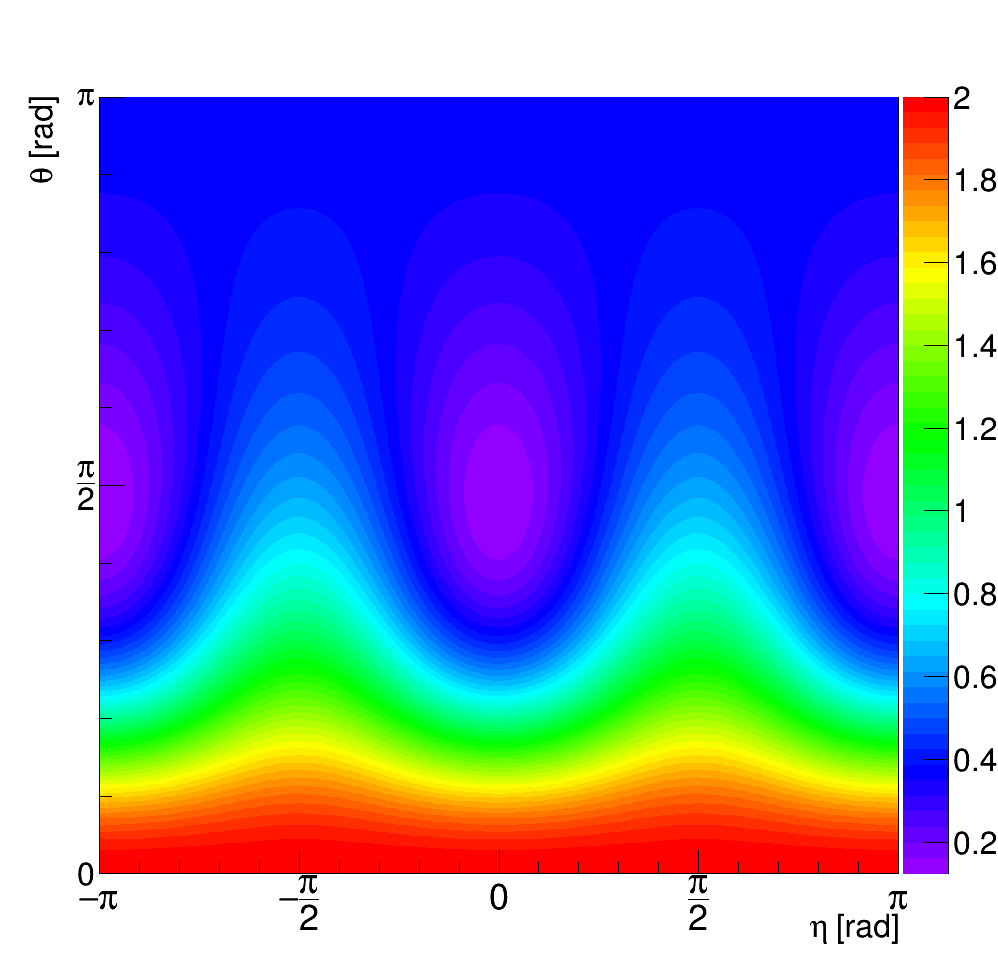}
\caption{Left: Cross section of the J-PET detector. Superimposed arrows indicate primary and scattered momenta of photons originating from para-positronium decaying in the center of the detector. Present J-PET tomograph is built from plastic scintillators strips arranged axially in three rings with radii of $425$~mm, $467.5$~mm and $575$~mm.
Right: Klein-Nishina differential cross section~(eq. \ref{eq:KNcross}) as a function of $\eta$ and $\theta$ angles (see text).
%
\label{fig:ComptonSchema}
}
\end{figure}

J-PET is a multipurpose detector designed for the development of medical imaging~\cite{NIM2014,NIM2015,PMB2016,Raczynski2017}, for studies of discrete symmetries in decays of positronium atoms~\cite{ACTA2016}, as well as for investigations of multipartite quantum entanglement of photons originating from positronium annihilation~\cite{Beatrix-Science-Report2017,Nowakowski}. J-PET is built from $192$ plastic scintillator strips arranged axially in three cylindrical layers~\cite{ACTA2017}. The cross section of the detector is shown in the left panel of Fig.~\ref{fig:ComptonSchema}. 
Photons from the $e^+e^-$ annihilation interact in plastic scintillators predominantly via the Compton effect and in the J-PET detector a few percent of them undergo secondary scatterings. Events with multiple scatterings may be used to estimate the linear polarization of the initial photon at the moment of its interaction. Taking into account that the scattering is most likely at an angle orthogonal to the polarization we may estimate polarization direction by~\cite{ACTA2016}: $\hat{\epsilon} = \hat{k} \times \hat{k'}$ where $\hat{k}$ and $\hat{k'}$ denotes the momentum versors of the  photon before and after the Compton scattering, respectively. 
Access to the polarization degree of freedom in the case of measurements of photons from the decays of positronium opens new perspectives for studies of the discrete symmetries~\cite{ACTA2016,Acin} and multipartite quantum entanglement~\cite{Beatrix-Science-Report2017,MUB}. 
In this article, we discuss the physical limits of the accuracy for the polarization determination of annihilation photons via Compton scattering.
\section{Compton scattering and Klein-Nishina formula}
Angular distribution of a scattered radiation of a linear polarized incoming photon can be described by the Klein-Nishina formula \cite{Klein2013,Evans1958}:
\begin{eqnarray}
\frac{d\sigma (E, \theta, \eta)}{d\Omega} &=& \frac{r_{0}^2}{2} \left(\frac{E^{\prime}}{E}\right)^2 \left(\frac{E}{E^{\prime}} + \frac{E^{\prime}}{E} - 2\sin^2{\theta}\cos^2{\eta}\right) 
\label{eq:KNcross}\\
&=&\frac{r_{0}^2}{2} \left(\frac{E^{\prime}}{E}\right)^2 \left(\frac{E}{E^{\prime}} + \frac{E^{\prime}}{E}-\sin^2\theta\right)\left\lbrace 1-\mathcal{V}(\theta,E)\cos(2\eta)\right\rbrace
\end{eqnarray}
with
\begin{eqnarray}
E^{\prime}(E, \theta) &=& \frac{E}{1 + \frac{E}{m_{e}c^2}(1 - cos{\theta})}\;,
\end{eqnarray}
where $E$ is the energy of initial photon, $E^{\prime}$ is the energy of photon after scattering, $\theta$ is the Compton scattering angle and $\eta$ is the angle between scattering and polarization planes. 
The expression $\mathcal{V}(\theta,E)$ quantifies the interference contrast, the \textit{a priori} visibility~\cite{HiesmayrComplementarity}. It is a typical quantity showing up in any double-slit-like scenario. Particularly, it has been shown of the Mott scattering, namely Rutherford scattering with identical particles and in the decay of neutral mesons, which are superpositions of particle and antiparticle states~\cite{HiesmayrComplementarity}. Obviously, if the visibility is close to zero, the oscillation due to the polarization degree of freedom are not observable. In this case no information on the polarization degree of freedom can be deduced. For $511$~keV photons the maximum of the visibility is obtained for a scattering angle of $\theta=81.67^{\circ}$ and the minimum of the visibility is obtained for small and large scattering angles $\theta$, independent of the energy. 

\begin{figure}[H]
\centering
  \includegraphics[width=0.49\textwidth]{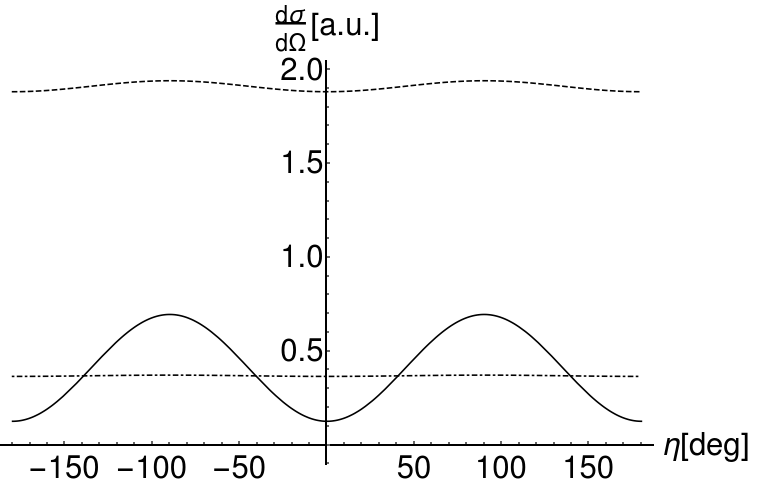}
  \includegraphics[width=0.49\textwidth]{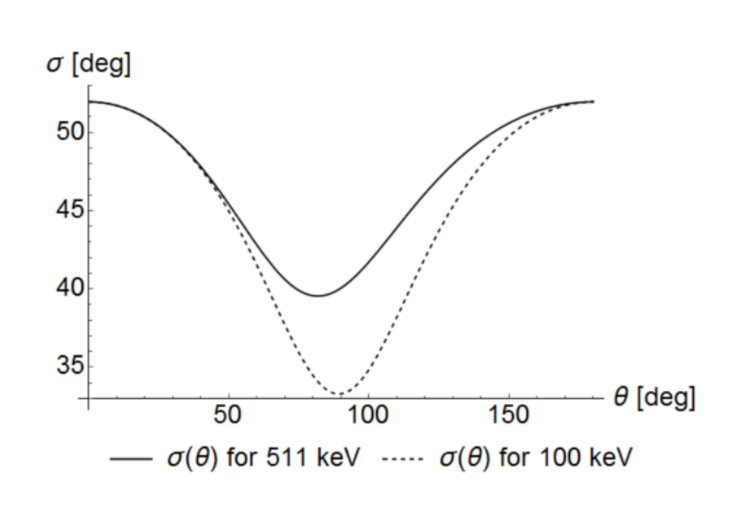}
\caption{
Left: Klein-Nishina differential cross section~(\ref{eq:KNcross}) for photons with initial energy $E = 511keV$ presented as a function of angle $\eta$. Dashed-, solid-, and dot-and-dashed-line show results for  scattering at angles  $\theta = 10^\circ$, $\theta = 81.67^\circ$, and $\theta = 170^\circ$, respectively.
Right: Uncertainty ($=$standard deviation) of the determination of the polarization direction for $511$keV photons (solid line) and $100$keV photons (dashed line) via Compton scattering~\cite{ACTA2016}. For definition of $\theta$ and $\eta$ angles see text.
\label{fig:KleinNishina}
}
\end{figure}

The right panel of Fig.~\ref{fig:ComptonSchema} shows the double differential cross section for Compton scattering of $511$~keV photons, as a function of $\eta$ and $\theta$ angles. As expected, for $\theta$ around $82^\circ$ a most pronounced modulation of the cross section as a function of the $\eta$ angle is observed. The amplitude of modulations decreases towards higher and lower values of $\theta$. A quantitative comparison of this dependence is shown for three chosen angles in the left panel of Fig.~\ref{fig:KleinNishina}.

The distributions shown in the left panel of Fig.~\ref{fig:KleinNishina} may be interpreted as a resolution function for the determination of the polarization direction by $\hat{\epsilon}$. The $\sigma$ values of these resolution functions obtained by the fit of the Gaussian in the scattering angle range between $0^\circ$ and $180^\circ$ is indicated for the $511$~keV photons by the solid line in the right panel of Fig.~\ref{fig:KleinNishina}. The dashed line shows the results for the energy of $100$~keV. It demonstrates that the lower the photon energy the better the direction of photon polarization may be estimated. This indicates that in the positronium decay into three photons, with photons energies less than $511$~keV the visibility improves with respect to the decay into two photons.

\section{Summary}
The J-PET tomograph built from plastic scintillators enables measurements of the polarization of photons at an event-by-event basis. In this article it was shown that the physical limitations for the accuracy of the estimation of the linear polarization direction of the 511~keV photons, due to the nature of the Compton scattering, is equal to $40^\circ$ ($\sigma$) for $\theta\approx 82^\circ$ and it worsens towards smaller and larger scatterings angles. As a result we have shown that studies of entanglement and discrete symmetries in positronium decays involving polarization should be concentrated for scattering angles around $82^\circ$ while for the forward and backward scatterings the information about polarization direction is not attainable. 

\begin{acknowledgements}
The authors acknowledge technical and administrative support of A. Heczko, M. Kajetanowicz and W. Migda\l{}. This work was supported by The Polish National Center for Research
and Development through grant \\ \mbox{INNOTECH-K1/IN1/64/159174/NCBR/12}, the
Foundation for Polish Science through the MPD and TEAM/2017-4/39 programmes, the National Science Centre of Poland through grants no.
\mbox{2016/21/B/ST2/01222}, \mbox{2017/25/N/NZ1/00861},
the Ministry for Science and Higher Education through grants no. \mbox{6673/IA/SP/2016},
\mbox{7150/E-338/SPUB/2017/1} and \\ \mbox{7150/E-338/M/2017}, and the Austrian Science Fund FWF-P26783.
\end{acknowledgements}



\end{document}